\documentstyle[12pt]{article}
\newcommand{\journal}[4]{{\em #1~}#2\,(19#3)\,#4;}

\newcommand{\pr}{\journal {Phys. Rev.}}

\newcommand{\prl}{\journal {Phys. Rev. Lett.}}

\newcommand{\cmp}{\journal {Comm. Math. Phys.}}

\newcommand{\np}{\journal {Nucl. Phys.}}
\newcommand{\pl}{\journal {Phys. Lett.}}

\newcommand{\prep}{\journal {Phys. Reports}}

\makeatletter
\@addtoreset{equation}{section}
\makeatother

\def\Lp{\displaystyle{\biggl(}}
\def\Rp{\displaystyle{\biggr)}}

\newcommand{\k}{\kappa}

 \renewcommand{\S}{\Sigma}

\newcommand{\DD}{{\cal D}}

\newcommand{\NN}{{\cal N}}

\newcommand{\SS}{{\cal S}}

\newcommand{\WW}{{\cal W}}

\newcommand{\ZZ}{{\cal Z}}

\newcommand{\complex}{{\kern .1em {\raise .47ex
\hbox {$\scriptscriptstyle |$}}
    \kern -.4em {\rm C}}}
\newcommand{\real}{{{\rm I} \kern -.19em {\rm R}}}
\newcommand{\rational}{{\kern .1em {\raise .47ex
\hbox{$\scripscriptstyle |$}}
    \kern -.35em {\rm Q}}}
\renewcommand{\natural}{{\vrule height 1.6ex width
.05em depth 0ex \kern -.35em {\rm N}}}

\newcommand{\pa}{\partial}
\newcommand{\pad}[2]{{\frac{\partial #1}{\partial #2}}}
\newcommand{\fud}[2]  {{\displaystyle{\frac{\delta #1}{\delta #2}}}}

\newcommand{\sla}{\raise.15ex\hbox{$/$}\kern -.57em}

\newcommand{\twiddle}{\lower.9ex\rlap{$\kern -.1em\scriptstyle\sim$}}

\newcommand{\equ}[1]{(\ref{#1})}

\newcommand{\eq}{\begin{equation}}
\newcommand{\eqn}[1]{\label{#1}\end{equation}}
\newcommand{\eea}{\end{eqnarray}}
\newcommand{\eqa}{\begin{eqnarray}}
\newcommand{\eqan}[1]{\label{#1}\end{eqnarray}}
\newcommand{\ba}{\begin{array}}
\newcommand{\ea}{\end{array}}
\newcommand{\eqac}{\begin{equation}\begin{array}{rcl}}
\newcommand{\eqacn}[1]{\end{array}\label{#1}\end{equation}}

\renewcommand{\pad}[2]{{\displaystyle{\frac{\partial #1}{\partial #2}}}}

\newcommand{\intx}{\int d^3 \! x \, }

\newcommand{\dphi}{\phi^\dagger}
\newcommand{\dgamma}{\gamma^\dagger}
\newcommand{\wA}{\widehat A}
\newcommand{\wphi}{\widehat \phi}
\newcommand{\wdphi}{{\widehat \phi}^{\dagger}}

\begin{document}
\def\ftoday{{\sl  \number\day \space\ifcase\month
\or Janvier\or F\'evrier\or Mars\or avril\or Mai
\or Juin\or Juillet\or Ao\^ut\or Septembre\or Octobre
\or Novembre \or D\'ecembre\fi
\space  \number\year}}


\vspace{8mm}
\setcounter{page}{1}

\titlepage

\begin{center}

{\huge Non-Renormalization Properties of the Chern-Simons Action
Coupled to Matter}

\vspace{1cm}

{\Large A. Blasi}\footnote{On leave of absence from Dipartimento di Fisica -
Universit\'a di Trento,
38050 Povo (Trento) - (Italy)}

\vspace{.3cm}
{\it Laboratoire d'Annecy-le-Vieux de Physique de Particules\\
Chemin de Bellevue BP 110, F - 74941 Annecy-le-Vieux Cedex, France}

\vspace{.5cm}

{\Large N. Maggiore}

\vspace{.3cm}

{\it
Dipartimento di Fisica -- Universit\`a di Genova\\
Istituto Nazionale di Fisica Nucleare -- sez. di Genova\\
Via Dodecaneso, 33 -- 16146 Genova (Italy)}

\vspace{.3cm}
and

\vspace{.3cm}

{\Large S. P. Sorella}\footnote{Supported in part by the
Swiss National Science Foundation}

\vspace{.3cm}
{\it D\'epartement de Physique Th\'eorique, Universit\'e de Gen\`eve\\
CH--1211 Gen\`eve 4, Switzerland}

\end{center}

\vspace{1.7cm}

\begin{center}
\bf ABSTRACT
\end{center}
{\it
We analyze an abelian gauge model in 3 dimensions which
includes massless
scalar matter fields. By controlling the trace anomalies with a local
dilatation
Ward identity, we show that, in perturbation theory and within the BPHZL
scheme, the Chern-Simons term has no radiative corrections. This implies,
in particular, the vanishing of the corresponding $\beta$ function
in the renormalization group equation. }

\vfill
GEF-Th-7/1992 \\
ENSLAPP-A-380-92 \hfill April 1992
\newpage
{\large     

\section{Introduction}
The topological nature of the Chern-Simons action in three space-time
dimensions leads naturally to the conjecture that it should be insensitive
to local deformations such as the ones induced by the perturbative
renormalization process of a local quantum field theory. That this is indeed
so for a pure, non abelian gauge field theory, was first confirmed by explicit
calculations~\cite{enore,birm} and later proven to all perturbative
orders~\cite{genova,ginevra}.

Although the techniques of the proofs adopted in this case are not directly
extendible to a gauge field interacting with matter, in particular when
scalar fields are present
\footnote{The case of only gauge and fermions fields can be treated
with a slight extension of the method of ref~\cite{dil}, indeed
it is well known that fermion fields without Yukawa couplings
are cohomologically trivial~\cite{fermions}.}, again the
explicit low order calculations
confirm that the Chern-Simons coupling constant has vanishing
$\beta$-function~\cite{sodano,kazakov}.

Here we propose a proof of this assertion, valid to all perturbative orders,
which is based on local dilatation invariance ( general coordinate
transformations in the Weyl's scheme ) as discussed in~\cite{dil}.
The idea of the
proof is that the trace anomalies affecting the local dilatation Ward identity
can be controlled, in the flat limit, by a single dimensionless scalar
external field $\sigma(x)$, whose infinite couplings are all identified by
those appearing in the classical $\sigma$-independent action.

Now the Chern-Simons term has the peculiarity of not being locally gauge
invariant, its variation being a total divergence. Therefore, in this scheme
where gauge and local dilatation invariance can be imposed independently,
it follows that this term does not generate $\sigma$-field
counterterms at any order and thus cannot contribute to the trace anomaly.

As expected this leads directly to the conclusion that the corresponding
$\beta$-function vanishes. Moreover, in the course of the proof we shall
also verify the stronger result that the Chern-Simons coupling does not
renormalize at all.

This is also an expected consequence of the topological nature of the coupling
which does not allow an identification in terms of a local normalization
condition; hence the only way out is, in a local quantum field theory
context, a complete insensitivity to radiative corrections.

For simplicity we shall treat the case of an abelian gauge field interacting
with a massless scalar field $\phi$ with a $(\dphi \phi)^3$ self-interaction.
The discussion proceeds within the $BPHZL$~\cite{zimm}
regularization independent scheme;
a more complete analysis including the Yang--Mills term as well as
massive scalar and fermion fields is
in preparation~\cite{prep}.

\section{The symmetries of the model}\label{section2}

The classical action in the Landau gauge is
\eq\ba{rl}
\S =
  \intx \Lp &\!\!
    {k \over 2} \varepsilon^{\mu\nu\rho} A_\mu \pa_\nu A_\rho
  - (D_\mu \phi)^{\dagger} (D^\mu \phi) - {\lambda \over 36} (\dphi \phi)^3 \\
         &\!\!
  +  \pa_\mu d A^\mu - \pa_\mu {\bar c} \pa^\mu c + i \dgamma \phi c
            -i \gamma \dphi c   \Rp  \ ,
\ea\eqn{classaction}
where
\eq
 D_\mu \phi = \pa_\mu \phi - i A_\mu \phi \ .
\eqn{covderiv}
The $BRS$ transformations are encoded in the nonlinear Slavnov identity
\eq
\SS({\S}) = \intx \Lp
    \pa_\mu c \fud{\S}{A_\mu} + d \fud{\S}{\bar c}
   + \fud{\S}{\dgamma}\fud{\S}{\phi}
   + \fud{\S}{\gamma}\fud{\S}{\dphi} \Rp = 0 \ .
\eqn{slavnov}
The action $\S$ \equ{classaction} satisfies the constraints:
\eq
 \fud{\S}{d} = -\pa A \ , \qquad \fud{\S}{\bar c} = \pa^2 c  \ ,
\eqn{gaugefix}
\eq
 \fud{\S}{c} = -\pa^2{\bar c} - i\dgamma\phi + i\gamma\dphi \ .
\eqn{ghosteq}

The renormalizations compatible with \equ{slavnov}, \equ{gaugefix} and
\equ{ghosteq} are:
\eq\ba{l}
 k \rightarrow Z_k k \\
 \lambda \rightarrow Z_\lambda \lambda \\
 \phi \rightarrow Z \phi \\
 \gamma^\dagger \rightarrow Z^{-1}\gamma^\dagger \ .
\ea\eqn{zeta}

Notice that the ghost fields as well as the lagrangian multipliers
do not renormalize due to the conditions \equ{gaugefix}, \equ{ghosteq} which
identify uniquely the gauge--fixing part of the action.

To include the local dilatations~\cite{dil} we have to introduce a metric
$g^{\mu\nu}(x) = \delta^{\mu\nu} + h^{\mu\nu}(x) $, a connection $\omega^\mu$
and the transformations laws with parameters $\lambda^\rho(x)$:
\eq\ba{l}
 \delta g^{\mu\nu} = \lambda^\rho \pa_\rho g^{\mu\nu}
   - g^{\rho\nu} \pa^\mu \lambda_\rho - g^{\rho\mu} \pa^\nu \lambda_\rho
   + {2 \over 3} g^{\mu\nu} \pa \lambda \\
 \delta \omega_\mu = \lambda^\rho \pa_\rho \omega_\mu
    + \omega_\rho \pa_\mu \lambda^\rho - {1 \over 6} \pa_\mu \pa\lambda \\
 \delta A_\mu = \lambda^\rho \pa_\rho A_\mu
    + A_\rho \pa_\mu \lambda^\rho \\
 \delta \phi = \lambda^\rho \pa_\rho \phi + {1 \over 6}\phi \pa\lambda \\
 \delta d    = \lambda^\rho \pa_\rho d    + {1 \over 3}d    \pa\lambda \\
 \delta {\bar c} = \lambda^\rho \pa_\rho {\bar c}
                  + {1 \over 3}{\bar c} \pa\lambda \\
 \delta c    = \lambda^\rho \pa_\rho c   \\
 \delta \gamma   = \lambda^\rho \pa_\rho \gamma
                  + {5 \over 6}\gamma  \pa\lambda \ .
\ea\eqn{diltransf}
The action \equ{classaction} becomes invariant under the above transformations
with the replacements
\eq\ba{l}
 \delta^{\mu\nu}  \rightarrow g^{\mu\nu}  \\
 \pa_\mu \phi \rightarrow ( \pa_\mu +   \omega_\mu )\phi \\
 \pa_\mu d    \rightarrow ( \pa_\mu + 2 \omega_\mu )d    \\
 \pa_\mu {\bar c}   \rightarrow ( \pa_\mu + 2 \omega_\mu ){\bar c}    \ .
\ea\eqn{replacements}
We will only be interested in the flat limit
$g^{\mu\nu} \rightarrow \delta^{\mu\nu}$, $\omega \rightarrow 0$ invariance
which we write as
\eq
{\Lp W_\rho (x) \S(\omega,g) \Rp}_{F.L} = 0  \ .
\eqn{Winvariance}
Due to the choice of the Weyl weights in \equ{diltransf} the $BRS$ and
$\delta$ symmetries commute so that they can be discussed independently.

As analyzed in detail in Ref.~\cite{dil}, the Weyl representation is unstable
under radiative corrections and the ensuing trace anomalies, in the flat
limit, can be reabsorbed by a single dimensionless external fields $\sigma(x)$
which is $BRS$ invariant and behaves as
\eq
\delta \sigma(x)    = \lambda^\rho \pa_\rho \sigma(x) +
              {\hbar \over 3}\pa\lambda \sigma(x)  \ .
\eqn{sigtransf}
The new Ward operator becomes
\eq
 \WW_\rho (x) = W_\rho(x)  + \pa_\rho \sigma(x)  {\fud{\ }{\sigma(x) }}
 - {\hbar \over 3}\pa_\rho {\fud{\ }{\sigma(x) }} \ ,
\eqn{Woperator}
which still commutes with the $BRS$ transformations.

The introduction of the $\sigma$ field forces us to consider the vertex
functional $\Gamma$ of the model as a double formal power series in $\sigma$
and $\hbar$ with coefficients $\Gamma^{(n,m)}$ at the order $\hbar^n \sigma^m$.
The infinite number of $\sigma$-couplings do not correspond however to
independent parameters; as shown in~\cite{dil} they are recursively
determined by
the local dilatation Ward identity in the flat limit in terms of the $\sigma$
independent ones.

Two remark are now in order; first and foremost, the Chern-Simons term cannot
couple to the $\sigma$ field since its $BRS$ variation is a divergence.
Second, the presence of the $\sigma$ field does not spoil the supplementary
conditions in \equ{gaugefix}, \equ{ghosteq} which acquire $\sigma$ dependent
terms on the right-hand side but which remain linear in the quantized fields.
We can therefore summarize the results thus far obtained by saying that in the
flat limit we can renormalize both the $BRS$ symmetry and the local dilation
Ward identity by allowing a $\sigma$ dependence for the $Z, Z_\lambda$
renormalization constants, but not for $Z_k$. The effective action,
in the sense of Zimmermann~\cite{zimm},  is
therefore
\eq\ba{rl}
\Gamma^{eff} =
  \intx \Lp &\!\!
    {k \over 2}Z_k \varepsilon^{\mu\nu\rho} A_\mu \pa_\nu A_\rho
   - (D_\mu Z(\sigma)\phi)^{\dagger} (D^\mu Z(\sigma)\phi)  \\
       &\!\!
   - {\lambda \over 36}( Z_\lambda(\sigma) {Z_\lambda(\sigma)}^{\dagger}
       \dphi \phi )^3   +  \pa_\mu d A^\mu -
   \pa_\mu {\bar c} \pa^\mu c \\
      &\!\!
    + i \dgamma \phi c
            -i \gamma \dphi c   \Rp  \ ,
\ea\eqn{gammaeff}
where the free parameters are identified by the normalization conditions:
\eq
 i \varepsilon_{\mu\nu\rho} { p^\rho \over p^2 }
\left. { \delta^2 \Gamma \over {\delta \wA_\mu(p)}{\delta A_\nu(0)}}
 \right|_{A=\phi=0,{\ }p^2=\mu^2} \equiv
\left.\Gamma_{A^2}(p^2) \right |_{p^2=\mu^2} = k \ ,
\eqn{cond1}
\eq
\left. { \delta^2 \Gamma \over {\delta \wdphi(p)} {\delta \phi(0)}}
\right|_{A=\phi=0,{\ }p^2=\mu^2} = - \mu^2 \ ,
\eqn{cond2}
\eq
\left.{ \delta^6 \Gamma \over
 {\delta \wdphi(p1)}{\delta \wdphi(p2)}{\delta \wdphi(p3)}
 {\delta \wphi(p4)}{\delta \wphi(p5)}{\delta \phi(0)} }
\right|_{A=\phi=0,{\ }symm.{\ }point {\ }\mu^2}  = - \lambda \ ,
\eqn{cond3}
where the hat denotes the Fourier transform.

\section{The renormalization group equation and the local dilatation
invariance}
Having identified the parameters in \equ{cond1}--\equ{cond3} we can write
immediately a renormalization group equation for the vertex functional
$\Gamma$ at $\sigma=0$; indeed a basis of Slavnov invariant differential
vertex operator is provided by
\eq
 \Lp \pad{\ }{k},\quad \pad{\ }{\lambda},\quad
   \NN=\intx \Lp \phi \fud{\ }{\phi} - \gamma^\dagger
   \fud{}{\gamma^\dagger}\Rp\ ,
\quad
\NN^\dagger=\intx \Lp \dphi \fud{\ }{\dphi} - \gamma
   \fud{}{\gamma}\Rp\Rp\ ,
\eqn{vertexop}
so that we have
\eq
\Lp \mu \pad{\ }{\mu} + \beta_k \pad{\ }{k} + \beta_\lambda \pad{\ }{\lambda}
+\gamma_\phi \Lp\NN + \NN^\dagger\Rp\Rp \Gamma = 0 \ ,
\eqn{groupequ}
where the coefficients $\beta_k, \beta_\lambda, \gamma_\phi$ can be computed
with the help of the normalization conditions \equ{cond1}--\equ{cond2}, in
particular we have
\eq
\left. \Lp \mu \pad{\Gamma_{A^2}(p^2)}{\mu} \Rp \right|_{p^2=\mu^2} =
-\beta_k \ .
\eqn{betakequ}
In order to establish a connection with the Ward identity
\eq
{\Lp \WW_\rho(x) \Gamma \Rp}_{Flat-limit} = 0 \ ,
\eqn{WWflatlim}
let us consider the insertion\footnote{This insertion is well defined if the
model has no infrared problems, which is also assumed in deriving the
renormalization group equation; otherwise we remain with the local dilation
Ward identity which controls the trace anomalies.}
\eq
\left. \pa_\sigma \Gamma(\sigma)\right|_{\sigma=0}
  \equiv \intx \left. \fud{\Gamma(\sigma)}{\sigma(x)} \right|_{\sigma=0} \ .
\eqn{sigmains}
Moreover, by integrating the Ward operator in \equ{Woperator} with parameter
$\lambda^\rho = x^\rho$ we obtain
\eq
\Lp \DD + \hbar \pa_\sigma \Rp \Gamma(\sigma) = 0 \ ,
\eqn{dilatequ}
where
\eq
\DD = \sum_{\psi} \intx \Lp x^\rho \pa_\rho\psi + d_\psi \Rp
         \fud{\ }{\psi} \ ,
\eqn{DDoperator}
and $\psi$ denotes collectively the $(A_\mu,{\ }\phi,{\ }\dphi)$ fields with
canonical dimension $d_\psi = \Lp 1\ ,\frac{1}{2}\ ,\frac{1}{2}\Rp$.

By considering \equ{dilatequ} at $\sigma=0$ we immediately find
\eq
 \mu \pad{\Gamma}{\mu} = \hbar \left.
\pa_{\sigma}\Gamma(\sigma) \right|_{\sigma=0} \ .
\eqn{muequ}
Let us now consider in detail the scaling behaviour imposed by \equ{dilatequ}
on the coupling $\Gamma_{A^2}(p^2)$ at the various orders in $\sigma$ and
$\hbar$, with the constraint that the effective action in \equ{gammaeff} does
not contain $\sigma$ field couplings with the Chern-Simons term.

In Fourier transform we obtain for $\Gamma_{A^2}^{(n,m)}(p^2)$ the recursive
relation
\eq
 p_\rho \pad{\Gamma_{A^2}^{(n,m)}}{p_\rho} + \Gamma_{A^2}^{(n-1,m+1)} = 0 \ .
\eqn{Gammarelat}

Recalling that $\Gamma^{eff(n,m)}_{A^2}=0$ for $m \ge 1$, and using
the Ward identity \equ{WWflatlim}, we shall be able to identify
the general solution of $\Gamma^{(n,0)}_{A^2}(p^2)$ as a function of $p^2$
parametrized with coefficients which are dependent on the dimensionless
constant $\k$ and on $\ln{\mu^2}$; then  we shall employ the normalization
condition \equ{cond1} and the power counting
in order to find the $\mu^2$ dependence of $Z_k$.

For $n=1$, $m=0$ we get
\eq
p_\rho \pad{\Gamma^{(1,0)}_{A^2}(p^2)}{p_\rho} = 0 \ ,
\eqn{recurs1}
hence
\eq
\Gamma^{(1,0)}_{A^2} = \alpha^1_0 \ ,
\eqn{result1}
where $\alpha^1_0$ is a $p^2$-independent constant.

At the next order $n=2$, $m=0$ we obtain
\eq
p_\rho \pad{\Gamma^{(2,0)}_{A^2}}{p_\rho}
   + \Gamma^{(1,1)}_{A^2} = 0 \ ,
\eqn{recurs2}
and also, for $n=1$, $m=1$
\eq
p_\rho \pad{\Gamma^{(1,1)}_{A^2}}{p_\rho} = 0 \ ,
\eqn{recurs3}
hence
\eq
\Gamma^{(1,1)}_{A^2}(p^2) = \alpha^1_1 \ , \qquad
\Gamma^{(2,0)}_{A^2}(p^2) = -{1 \over 2}\alpha^1_1 \ln(p^2) + \alpha^2_0 \ .
\eqn{result2}
Proceeding inductively we find
\eq
\Gamma^{(n,0)}_{A^2}(p^2) = \sum_{j=0}^{n-1} {(-{1 \over 2})}^j
   \alpha^{n-j}_j (\ln{p^2})^j \ , \qquad n \ge 1 \
\eqn{result3}
where the $\alpha^{n-j}_j$ are functions of $\ln{\mu^2}$ and $\k$.
Now the power counting properties of $\Gamma^{(n,0)}_{A^2}(p^2)$ imply that
the above expression can be rewritten as
\eq
\Gamma^{(n,0)}_{A^2}(p^2) = \sum_{j=0}^{n-1}
   \gamma^n_j (\ln{p^2 \over \mu^2})^j \ , \qquad n \ge 1 \
\eqn{result4}
where the $\gamma^n_j$ are $\mu^2$ independent and uniquely determine the
$\ln{\mu^2}$ dependence of the $\alpha^{n-j}_j$ by
\eq
(-{1 \over 2})^j
   \alpha^{n-j}_j = \sum_{r=j}^{n-1}\gamma^n_r (-\ln{\mu^2})^{r-j}
\frac{j!}{(r-j)!r!} \ , \qquad n \ge 1 \ .
\eqn{}
The normalization condition \equ{cond1} yields
\eq
\gamma^n_0 = 0 \ , \qquad n \ge 1 \ ,
\eqn{}
so that the perturbative power series for the  $\beta_k$ and the
$Z_k^{(n)}$ renormalization constant are given , in terms of \equ{result4} by
\eq
\beta_k^{(1)} = 0\ ,\ \beta_k^{(n)} = \gamma_1^n \ ,\ Z_k^{(1)} = 0\  ,\
Z_k^{(n)} = \sum_{j=1}^{n-1}
   \gamma^n_j (-\ln{\mu^2})^j \ , \quad n \ge 2 \ .
\eqn{Zetakequ}
In particular we see that the one loop $\beta_k$ function is identically
zero and that the vanishing of the higher orders is related to the
vanishing of the $\gamma^n_1$ coefficients.

\section{Non-renormalization of the Chern-Simons term}

In order to prove that the $Z_k^{(n)}$'s vanish or equivalently that all
$\gamma^n_k$ also do so, we employ relations \equ{sigmains}, \equ{Zetakequ} and
the Quantum Action Principle~\cite{qap}.

Now all the Slavnov invariant, $\sigma$ dependent insertions are deducible
from the effective action in \equ{gammaeff}; expanding the renormalization
constants $Z_\lambda$, $Z$ to first order in $\sigma$
\eq\ba{rl}
Z_\lambda(\sigma) =& Z_\lambda^{(0)} + \sigma Z_\lambda^{(1)} + O(\sigma^2) \\
Z(\sigma) =& Z^{(0)} + \sigma Z^{(1)} + O(\sigma^2) \ ,
\ea\eqn{Zexpansion}
and considering the basis of insertions
\eq\ba{rl}
\Delta_{CS} =& {1 \over 2}\intx \varepsilon^{\mu\nu\rho}  A_\mu \pa_\nu A_\rho
\ , \\
\Delta_\lambda =& - {1 \over 36} \intx (\dphi \phi)^3 \ , \\
\Delta   =& - \intx {(D_\mu \phi)}^{\dagger}{(D^\mu \phi)} \ ,
\ea\eqn{basis}
we find
\eqa
&&\left. \Lp \mu  \pad{\ZZ}{\mu} \Rp \right |_{\sigma=0} =
  {1 \over \hbar}
 \left. \Lp \mu \pad{\Gamma^{eff}}{\mu} \cdot\ZZ \Rp \right |_{\sigma=0}
\label{ZZequ1}\\
&& =  \left. \Lp k\mu \pad{Z_k}{\mu} \Delta_{CS}
    + \mu {\pa (Z^{(0)}{Z^{(0)\dagger}}) \over \pa \mu} \Delta
 + \mu
{\pa {\Lp Z_\lambda^{(0)}(Z^{(0)}{Z^{(0)\dagger}})^3 \Rp} \over \pa\mu}
   \Delta_\lambda \Rp \cdot \ZZ \right |_{\sigma=0} \ ,\nonumber
\eea
and likewise
\eqa
&&\left. \Lp \hbar \pad{\ZZ}{\sigma} \Rp \right |_{\sigma=0} =
  \left. \Lp  \pad{\Gamma^{eff}}{\sigma}\cdot\ZZ \Rp \right |_{\sigma=0}
\label{ZZequ2}\\
&& =  \left. \Lp (Z^{(1)} + {Z^{(1)\dagger}})\Delta +
         \Lp 3 Z_\lambda^{(0)}(Z^{(1)} + {Z^{(1)\dagger}}) +
         Z_\lambda^{(1)} ( Z^{(0)}{Z^{(0)\dagger}} )^3 \Rp \Delta_\lambda
 \Rp \cdot \ZZ \right |_{\sigma=0} \ , \nonumber
\eea
where $\ZZ$ denotes the disconnected functional whose connected part
$\ZZ_c$ is given by the Legendre transform of the the vertex functional
$\Gamma$:
\eq
\ZZ_c(J,\sigma) = \Gamma(\sigma){\ }
  + \sum_{\varphi=all{\ }fields} \intx J\varphi \ .
\eqn{ZZetadef}
Since the left hand sides of \equ{ZZequ1}, \equ{ZZequ2} coincide, and we have
expandend on an independent basis, we obtain the relation
\eq
\mu \pad{Z_k}{\mu} = 0 \ ,
\eqn{finresult}
which, substituing \equ{Zetakequ}, implies recursively
\eq
\gamma^n_j = 0 \ , \qquad n \ge 1 \ .
\eqn{alphafinal}

The proof presented here proceeds
straightforwardly once the main point, {\it i.e.} the lack of local gauge
invariance of the
Chern-Simons term is recognized as the property which accounts for the absence
of the corresponding trace anomalies. Clearly a similar method can be applied
to discuss generalizations of this model,  for instance the inclusions
of massive scalars, fermions with Yukawa coupling etc.

\vspace{1cm}

\noindent{\large{\bf Acknowledgments}}: We would like to thank
O. Piguet for discussions on the subject of this paper;
one of the authors (A.B.) would like
to thank the ENSLAPP of Annecy-le-Vieux for hospitality.



\end{document}